\documentclass[
aip,
 amsmath,amssymb,
 reprint,%
]{revtex4-1}

\usepackage{graphicx}
\usepackage{dcolumn}
\usepackage{bm}

\usepackage[utf8]{inputenc}
\usepackage[T1]{fontenc}
\usepackage{mathptmx}

\usepackage{amsmath,amssymb}
\usepackage{bm}
\usepackage{color}

\newcommand{\INS}[1]{#1}

\begin{document}

\title{Noise stability of synchronization and optimal network structures}
\author{Yuriko Katoh}
\affiliation{NTT DATA Mathematical Systems Inc., Tokyo 160-0016, Japan}
\affiliation{Department of Information Sciences, Ochanomizu University, Tokyo 112-8610, Japan}
\author{Hiroshi Kori}
\email[Corresponding author: ]{kori@k.u-tokyo.ac.jp}
\affiliation{Department of Information Sciences, Ochanomizu University, Tokyo 112-8610, Japan}
\affiliation{Department of Complexity Science and Engineering, The
University of Tokyo, Chiba 277-8561, Japan}
\date{\today}

\begin{abstract}
 We provide a theoretical framework for quantifying the expected level of
 synchronization in a network of noisy oscillators.
 Through linearization around the synchronized state,
 we derive the following quantities as functions of the eigenvalues
 and eigenfunctions of the network Laplacian
 using a standard technique for dealing with multivariate Ornstein-Uhlenbeck processes:
 the magnitude of the fluctuations around a synchronized state
 and the disturbance coefficients $\alpha_i$ that represent 
 how strongly node $i$ disturbs the synchronization. With this approach, we can quantify the effect of individual nodes and links
on synchronization.
 Our theory can thus be utilized to find the optimal network structure for accomplishing the best synchronization.
 Furthermore, when the noise levels of the oscillators are heterogeneous, 
 we can also find optimal oscillator configurations,
 i.e., where to place oscillators in a given network depending on their
 noise levels.  We apply our theory to several example networks to
 elucidate optimal network structures and oscillator configurations.
 \end{abstract}


\def\del{\partial}

\maketitle

\begin{quotation}
Synchronization of rhythmic elements is essential in many systems.
To function properly and well,
rhythmic elements are required to maintain an appropriate 
synchronization pattern precisely.
What is the best network structure for accomplishing the best
synchronization? In other word, which elements should each element have
look at? Here, we develop a measure to quantify the precision of
synchronization for a given network. Using this measure,
we can quantitatively compare 
the stability of different networks and find the optimal network
structure. We can also determine where reliable or unreliable elements
should be placed in a given network.
\end{quotation}

\section{Introduction}
Synchronization of rhythmic elements, or oscillators, is ubiquitous and
 underlies various important functions \cite{winfree01,kuramoto84,pikovsky01}.
 For example, biological rhythms,
 including circadian rhythms and heartbeats,
 are generated by a population of cells acting periodically and
 synchronously \cite{winfree01,glass01}. 
Synchronization also plays a vital role in locomotion \cite{hoyt81,taga91,ijspeert08}. 
For each gait, the limbs perform rhythmic movements and maintain
a certain synchronization pattern. 
Synchronization is also essential in 
various artistic performances, including those by orchestras, choruses, and dancers \cite{stoklasa12,miyashita11,wuyts95}. 

In any example, to function properly and well, 
a population of oscillators is required to maintain an appropriate 
synchronization pattern, such as perfect synchrony, wave-like patterns, 
or more complex patterns. 
However, oscillators are inevitably exposed to noise. 
For example, the activity of a cell involves fluctuations due to 
various types of intrinsic and extrinsic noises \cite{elowitz02,faisal08}. Limbs 
experience perturbations from the ground or the surrounding fluid. 
Humans are unable to generate 
perfectly rhythmic actions, even in the absence of external disturbances. 
Such randomness disturbs synchronization and 
may hamper performance. 
Synchronization patterns must therefore be highly stable against the noise 
affecting individual oscillators.
Since synchronization occurs because of the interactions between the oscillators,
the structure of the interaction network is expected to strongly influence the
 synchronization stability.


The local stability problem of synchronous states is generally reduced to 
an eigenvalue problem of a particular class of stability matrices, which is often 
referred to as a network Laplacian $L$ or a Kirchhoff matrix \cite{arenas08}.
This class of matrices appears in
a variety of dynamical processes on networks and lattices,
such as random walks \cite{masuda17}, consensus problems \cite{olfati-saber07}, and
reaction-diffusion on networks \cite{nakao10}.
Consequently, there is a long history of studies of network 
Laplacians. In particular, the properties of the eigenvalues, or the
spectrum of the network Laplacians, have been studied intensively \cite{mohar1991laplacian,chung1997spectral}. 
The smallest non-zero eigenvalue of $L$, termed $\lambda_2$ in
this paper, often attracts attention because
its inverse provides a typical timescale that facilitates relaxation to
a synchronized state \cite{arenas08}.
It also provides a condition for the change of stability caused by variations in the system
 parameters, including changes in the network structure \cite{arenas08}.
For the synchronization of chaotic oscillators, the ratio of the
smallest to the largest eigenvalues,
 $\lambda_2 / \lambda_N$,
 also plays an important role in determining
 the stability of the network \cite{barahona2002synchronization}, and the 
optimal network structure that minimizes this ratio has been investigated
\cite{nishikawa2006synchronization}.

However, when we are concerned with the extent to which the synchronization 
pattern is precisely maintained in a network of noisy oscillators, 
knowledge of just a few dynamical modes is not 
sufficient, because every dynamical mode is excited at every time by 
noise. Therefore, we provide a theoretical framework here for quantifying the magnitude of 
the fluctuations around a synchronous state. 
Our framework is based on phase models, which describe oscillator 
networks to a good approximation when the coupling and noise are
sufficiently weak.
We are particularly interested in the case 
in which oscillators have different noise
strengths, because individual cells and humans experience different
noise levels.
We derive an expression for the magnitude of the fluctuations in an entire 
network as the weighted sum of the noise intensities of individual 
oscillators. This weight, termed the \lq\lq disturbance coefficient\rq\rq\ of 
a node, describes the extent to which an oscillator placed at that node 
disturbs the synchronization of the network. The disturbance coefficients of a network depend 
on the network structure, which may differ significantly among the nodes. 
Our theory can thus be utilized to find an optimal network 
structure that minimizes the fluctuation level and to find 
an optimal oscillator configuration; i.e., to determine at which nodes oscillators with 
higher or lower noise strengths should be placed in a given network.

\section{Theory} \label{sec:theory}
We first present our theoretical framework; we outline our theory before going into detail about it.
In Sec.~\ref{sec:synchronization},
We begin by considering a particular class of phase models that describe 
the networks of $N$ interacting oscillators admitting perfect synchrony (i.e., an in-phase state) in the absence of noise.
The level of synchronization can be characterized by the Kuramoto 
order parameter $r(t)$ $(0\le r \le 1)$, which assumes $r=1$ in the absence of noise and 
typically decreases as the strength of the noise increases. 
We are concerned with the expectation (i.e., the ensemble average) of $r$ 
for a given network and noise strength. In Sec.~\ref{sec:linearization},
we derive an expression for this quantity, denoted by $Q$, 
by assuming weak noise and linearizing 
the system around the in-phase state.
The problem with which we are concerned is then reduced to 
a general class of linear dynamical systems, which are described by a network
Laplacian $L$. 
We derive $Q$ as a function of the eigenvalues and eigenvectors of $L$ 
and of the individual noise strengths $\eta_i$ ($1\le i \le N$). 
In the derivation, we assume $L$ is diagonalizable; however, 
we also propose a method to treat a 
non-diagonalizable Laplacian $L$ (Sec.~\ref{sec:nondiagonalizable}).
In Sec.~\ref{sec:generalization}, we show that our theory can also be 
applied to a more general class of phase models and synchronized 
states. 

Examples and numerical verification follow in Secs.~\ref{sec:examples} and 
~\ref{sec:verification}, respectively. 

\subsection{Synchronization of oscillator networks} \label{sec:synchronization} 
We consider a network of self-sustained oscillators 
that are subjected to independent noise. When the coupling and noise are weak, the system is 
described by a phase model to a good approximation \cite{winfree67,kuramoto84}. By further assuming that all the oscillators are identical, 
it is appropriate to consider the system 
\begin{align}
\dot {\phi_i} (t) = \omega +
 \sum_{j=1}^N A_{ij} f(\phi_j - \phi_i) + \xi_i (t),
  \label{pm}
\end{align}
where $\phi_i$ $(1 \leq i \leq N)$ is the phase of the $i$th oscillator, 
$\omega$ is the natural frequency, 
$A_{ij} \ge 0$ is the weight of a directed edge that 
describes the strength of the coupling from the $j$th oscillator to the $i$th oscillator, 
$f$ is a $2\pi$-periodic function, 
and the $\xi_i$ represents independent Gaussian white noise. The latter variables satisfy 
\begin{align}
 \langle \xi_i (t) \rangle = 0, \quad
 \langle \xi_i (t) \xi_j (s) \rangle
 = \eta_i \delta_{ij} \delta (t - s),
 \label{noise}
\end{align}
where $\langle \cdot \rangle$ represents the expectation value 
and $\eta_i \ge 0$ is the strength of the noise to which the $i$th oscillator is subjected. 
We assume $f(0)=0$ and $f'(0)>0$. The former implies 
that the coupling vanishes when all the 
oscillators are in phase; i.e., $\phi_i=\phi_j$ for all $i$ and $j$. 
The latter implies that the in-phase state of two mutually coupled 
oscillators is linearly stable in the absence of noise. 
This type of coupling typically arises in chemical and biological oscillators coupled 
electrically or diffusively 
\cite{kopell2004chemical,kiss05,miyazaki06,kori14,stankovski2017coupling}.
We set $f'(0)=1$ without loss of generality.
Our theory may be generalized to more general phase models, as described 
in Sec.~\ref{sec:generalization}.

In this setting, our oscillator network has an in-phase state (i.e., 
the completely synchronized state), which is given 
by
\begin{equation}
 \phi_i = \omega t + C,
\end{equation}
where $C$ is an arbitrary constant.
We assume that this state is stable, which holds true under mild
conditions, as detailed in Sec.~\ref{sec:linearization}.
We also assume that the noise is sufficiently weak 
so that the system fluctuates weakly around the in-phase state. 
We are concerned with the magnitude of the fluctuations of this system. 

To quantify the level of synchronization, we introduce the Kuramoto order
parameter $r$ $(0\le r \le 1)$, defined as
\begin{align}
 r e^{i \theta} =
 \frac{1}{N} \sum_{j=1}^{N} e^{i \phi_j},
 \label{kuramoto}
\end{align}
where $\theta$ can be interpreted as the mean phase of the oscillators. 
When the system is nearly in-phase, $\phi_j - \theta$ is small. 
By rewriting Eq.~\eqref{kuramoto} as $r = \frac{1}{N} \sum_{j=1}^{N} e^{i
(\phi_j-\theta)}$ and dropping the terms of $O \left[ (\phi_j - \theta)^3 \right]$,
we obtain
\begin{align}
 r
 &= \frac{1}{N} \sum_{j=1}^{N}
 \left(1 - \frac{(\phi_j - \theta)^2}{2}
 + i (\phi_j - \theta) \right).
\end{align}
By equating the imaginary parts of both sides, we find 
\begin{equation}
 \theta = \frac{1}{N} \sum_{j=1}^N \phi_j.
\end{equation}
By equating the real parts of both sides and introducing $x_i=\phi_i - \omega t$, we obtain 
\begin{align}
 r = \frac{1}{N} \sum_{j=1}^{N}
 \left[ 1 - \frac{(x_j - \overline x)^2}{2} \right],
\end{align}
where
\begin{align}
 \overline x = \frac{1}{N} \sum_{j=1}^N x_j.
\end{align}
The expectation value of $r$ is thus given by 
\begin{align}
 \langle r \rangle = 1 - \frac{Q}{2},
 \label{r_expectation}
\end{align}
where
\begin{align}
 Q = \frac{1}{N} \sum_{j=1}^{N} \langle  (x_j - \overline x)^2 \rangle.
 \label{Q_def}
\end{align}
The quantity $Q$ can be interpreted as the variance of the phases $\phi_i$ 
when the system is nearly in phase. 
The smaller the value of $Q$, 
the better the system is synchronized. 
Below, based on linearization and diagonalization of our model, 
we derive an expression for $Q$. 

\subsection{Linearized system}\label{sec:linearization}
We linearize Eq.~\eqref{pm} for
small phase differences $\phi_j - \phi_i$ ($1 \le i,j \le N$) and
substitute $\phi_i = \omega t + x_i$ to obtain
\begin{align}
 \dot{x_i} = \sum_{j=1}^N A_{ij} (x_j - x_i) + \xi_i,
 \label{model_x}
\end{align}
or
\begin{align}
 \dot{\bm x} = - L \bm x+ \bm \xi, 
 \label{model}
\end{align}
where $\bm x=(x_1, \ldots, x_N)^{\rm T}$ and $\bm \xi=(\xi_1, \ldots,
\xi_N)^{\rm T}$, and the network Laplacian $L=(L_{ij})$ is given by
\begin{align}
 L_{ij} =
 \begin{cases}
  - A_{ij} & \mathrm{for} \, i \neq j, \\
  \displaystyle \sum_{i' \neq i} A_{ii'} & \mathrm{for} \, i = j.
 \end{cases}
\end{align}
Equation \eqref{model} is a particular class of multivariate
Ornstein-Uhlenbeck processes. When $L$ is diagonalizable, which we
assume below, many quantities can be derived analytically \cite{Riskenbook}.
We denote the eigenvalues of $L$ by $\lambda_n$ $(1 \leq n \leq N)$ and 
their corresponding right and left eigenvectors by 
$\bm {u} ^{(n)} = \left(u_1^{(n)}, u_2^{(n)}, \ldots, u_N^{(n)}
\right)^{\rm T}$
and $\bm {v} ^{(n)} = \left(v_1^{(n)} \; v_2^{(n)} \; \cdots \; v_N^{(n)} \right)$, respectively; i.e., 
\begin{align}
 L \bm {u} ^{(n)} &= \lambda_n \bm {u} ^{(n)}, \label{u} \\
 \bm {v} ^{(n)}  L &= \lambda_n \bm {v} ^{(n)}. \label{v}
\end{align}
Note that $\bm {u}^{(n)}$ and $\bm {v}^{(n)}$ are column and row
vectors, respectively.
Because $L$ is assumed to be diagonalizable, 
these eigenvectors can be chosen to be bi-orthonormal; i.e., 
\begin{equation}
 \bm {v} ^{(m)} \bm {u} ^{(n)} = \delta _{mn}. \label{orthonormal}
\end{equation}
For a symmetric matrix $L$, the right and left eigenvectors are 
parallel to each other; thus, we set $\bm {v} ^{(m)} = \bm {u} ^{(m){\rm T}}$ and 
normalize the eigenvectors as $\bm {u}^{(m)}\cdot \bm {u} ^{(n)} = \delta _{mn}$.

One of the eigenvalues of $L$ is zero; it is denoted by $\lambda_1 = 0$, and 
its corresponding right eigenvector is denoted by 
\begin{align}
  \bm {u} ^{(1)} = 
  (1, 1, \cdots, 1)^{\rm T}.
  \label{u1}
\end{align}
When the in-phase state is stable, 
we have
\begin{align}
 0 = \lambda_1 < \mathrm{Re}\, \lambda_2 \leq \mathrm{Re}\, \lambda_3
  \leq \hdots \leq \mathrm{Re} \, \lambda_N,
  \label{eigenvalue}
\end{align}
where $\mathrm{Re} \, \lambda$ denotes the real part of $\lambda$.
When $A_{ij} \ge 0$ for $1\le i,j \le N$,
Eq.~\eqref{eigenvalue} holds true under the following mild condition:
all the nodes are reachable from a single node
along directed paths, where the directed path from node $j$ to $i$ is
assumed to be present when $A_{ij}>0$ \cite{ermentrout92,arenas08}. Strongly connected networks suffice this condition.

By diagonalizing 
Eq.~\eqref{model} using the eigenvectors defined above, 
we can solve Eq.~\eqref{model} to derive the expression for $Q$ given in 
Eq.~\eqref{Q_def}. As shown in detail in Appendix~\ref{sec:derivation}, we obtain 
\begin{subequations}
 \begin{align}
  Q &= \sum_{i=1}^N \alpha_i \eta_i, \label{Q}\\
  \alpha_i &=
  \sum _{m,n = 2}^{N}
  \frac{
  \overline {u ^{(m)}  u ^{(n)}} - \overline {u ^{(m)}} \ \overline {u
  ^{(n)}}}
  {\lambda_m + \lambda_n}
  v_{i}^{(m)} v_{i}^{(n)}.
  \label{alpha}
 \end{align} \label{Q_alpha}%
\end{subequations}
where $\overline {u ^{(m)}}= \frac{1}{N} \sum_{i=1}^N u_i^{(m)}$, and
$\overline {u^{(m)}u^{(n)}}= \frac{1}{N} \sum_{i=1}^N u_i^{(m)}u_i^{(n)}$.
Thus, as given in Eq.~\eqref{Q_alpha}, 
fluctuations around the synchronous state are expressed as 
the summation of individual noise 
strengths $\eta_i$, each weighted by $\alpha_i$, which we call the 
disturbance coefficient of a node $i$. 
Oscillators placed at the nodes with larger values of $\alpha_i$ tend to 
disturb the synchronization more strongly.

For a symmetric matrix $L$, 
Eq.~\eqref{alpha} reduces to (see Appendix~\ref{sec:derivation}) 
\begin{align}
 \alpha_i = \frac {1}{2N} \sum_{n = 2}^{N}
 \frac {\left(u_{i}^{(n)} \right) ^2 }{\lambda_n}.
 \label{alpha_sym}
\end{align}
Further, by assuming homogeneous noise strengths, i.e., $\eta_i = \eta$, 
Eq.~\eqref{Q} reduces to
\begin{align}
 Q = \frac{\eta}{2N}\sum_{n=2}^N  \frac {1}{\lambda_n}.
 \label{Q_sym}
\end{align}
Equation \eqref{Q_sym} has already been derived in Ref. \cite{yanagita14}, which 
focuses on symmetric Laplacians $L$ and homogeneous noise strengths.

\subsection{The non-diagonalizable case} \label{sec:nondiagonalizable}
Our derivation above was based on the assumption that $L$ is diagonalizable. 
However, we may also be interested in networks that yield 
non-diagonalizable matrices $L$, which we consider in Sec.~\ref{sec:34}. 
Even when $L$ is non-diagonalizable, we may obtain values for $Q$ and $\alpha_i$ 
in the following manner. 

We assume that we have a non-diagonalizable Laplacian $L$. 
Then, we introduce $M$ extra parameters $\bm p=(p_1, p_2, \ldots, 
p_M) \in \mathbb R^M$ and add $p_k$ to $L_{i_k j_k}$ ($1\le k \le M, 1 \le i_k \le N,  1 \le j_k \le N$).
We denote the resulting matrix by $L(\bm p)$. By construction, we have 
$L=L(\bm 0)$. We may obtain a diagonalizable matrix $L(\bm p)$ if 
$M$ is sufficiently large and an appropriate set 
$\{(i_k,j_k)\}$ is chosen. We denote the resulting expression for $Q$ 
for $L(\bm p)$ by $Q(\bm p)$.
We may expect 
$Q(\bm 0)$ to describe the $Q$ value for the non-diagonalizable $L(\bm 0)$. 

We show that this method indeed works 
for the network considered in Sec.~\ref{sec:34}, 
which we verify numerically in Sec.~\ref{sec:verification}.

\subsection{Generalization} \label{sec:generalization}
In Sec.~\ref{sec:synchronization}, 
we considered a particular class of phase models, 
represented by Eq.~\eqref{model}, in order to consider a stable in-phase state. 
Our theory can also be extended to a more general class of phase models 
in which a stable phase-locked state exists. 
Important examples include phase waves and spirals in spatially 
extended systems \cite{ermentrout92,mkk10}.

We consider
\begin{align}
\dot {\phi_i} (t) = \omega_i +
 \sum_{j=1}^N B_{ij} f_{ij}(\phi_j - \phi_i) + \xi_i (t),
  \label{pm2}
\end{align}
where $\omega_i$ is the natural frequency of oscillator $i$, 
$B=(B_{ij})$ is the adjacency matrix, and 
$f_{ij}$ is a $2\pi$-periodic function that describes the coupling from
oscillator $j$ to oscillator $i$.
We assume that in the absence of noise, 
Eq.~\eqref{pm} has a phase-locked state 
\begin{equation}
 \phi_i(t) = \Omega t + \psi_i^*
  \label{phase_lock}
\end{equation}
for $1\le i \le N$. Here, $\Omega$ is the frequency of the synchronized 
state and the $\psi_i^*$ are constant phase offsets, which are found as 
solutions to the following set of equations: $\omega_i + \sum_{j=1}^N B_{ij}
f_{ij}(\psi_j^* - \psi_i^*)=\Omega$ ($1\le i \le N$).
Then, introducing $x_i(t) = \phi_i(t) - \Omega t - \psi_i^*$ and 
linearizing Eq.~\eqref{pm} for small $x_j - x_i$, 
we obtain exactly the same linear model as given by Eq.~\eqref{model}, 
where now 
\begin{equation}
 A_{ij} = B_{ij} f_{ij}'(\psi_j^* - \psi_i^*).
  \label{Aij_new}
\end{equation}
For such a phase-locked state, the magnitude of the fluctuations around 
the synchronized state can be quantified by Eq.~\eqref{Q_def}. 
Therefore, the theory presented in Sec.~\ref{sec:linearization} does not 
require any modification. Only the interpretation of $A_{ij}$ is 
slightly changed, as indicated in Eq.~\eqref{Aij_new}.

\section{Examples} \label{sec:examples}
Utilizing our theory, we now look for optimal network structures 
for several types of networks under various constraints. 
We assume that each oscillator has its own inherent noise 
strength and that we are allowed to place an oscillator at an arbitrary node in the network 
to make $Q$ as small as 
possible; i.e., we also consider the optimal configuration of oscillators. 

\subsection{Two nodes with two weighted edges} \label{sec:22}
\begin{figure}[t] 
 \includegraphics[width=4cm]{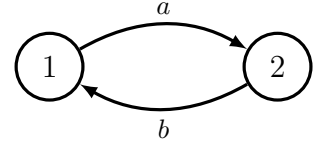}
 \caption{Network of two nodes and two edges, in which $Q$ is inversely 
 proportional to $a+b$ and does not depend on the ratio of $a$ to
 $b$.}
 \label{fig:two_nodes}
\end{figure}

We first consider a very simple network; i.e., two nodes with two 
weighted edges (Fig.~\ref{fig:two_nodes}). The corresponding Laplacian is 
\begin{align}
  L = \left(
  \scalebox{1.0}{$\displaystyle
  \begin{array}{cc}
   b & - b \\
   - a & a
  \end{array}
  $} \right),
\end{align}
which has the eigenvalues $\lambda_1 = 0$ and $ \lambda_2 = a + b$. 
Thus, the stability condition holds true when $a+b > 0$. 
The corresponding right and left eigenvectors are
\begin{align}
&  \bm u ^{(1)} = \left(1,1\right)^{\rm T},
 \bm u ^{(2)} = \left(- \frac{b}{a}, 1 \right)^{\rm T},\\
 &  \bm v ^{(1)}
  = \left(\frac {a}{b} \quad 1 \right)
,
  \bm v ^{(2)}
  = \left(- 1 \quad 1 \right).
\end{align}
Substituting these expressions into Eq.~\eqref{Q_alpha}, we obtain 
\begin{align}
 Q = \frac{ \eta_1 + \eta_2 } {16 (a+b) }.
 \label{Q_2nodes}
\end{align}
Here, $Q$ decreases with increasing $a+b$, in accordance with 
the behavior of the eigenvalues and is independent of the ratio of 
$a$ to $b$; i.e., there is no network-structure dependence in this 
particular example. 
Moreover, the disturbance coefficients $\alpha_1$ and $\alpha_2$ are identical, so $Q$ is independent of the oscillator configuration.

\subsection{Three nodes with three weighted edges} \label{sec:33}
\begin{figure}[t] 
 \includegraphics[width=8cm]{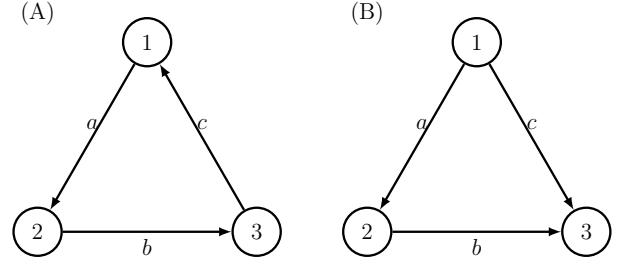}
 \caption{Networks of three nodes and three edges. (A) Feedback
 network. (B) Feedfoward network. The optimal weight distribution
 under the constraint $a+b+c=1$ and $\eta_i=\eta$ ($i=1,2,3$)
 is (A) $a=b=c=\frac{1}{3}$ and (B) $a=c=\frac{1}{2}$, $b=0$. The
 corresponding $Q$ value is $\frac{\eta}{3}$ for both networks; these two optimal
 networks are equivalently noise-tolerant.}
  \label{fig:33}
\end{figure}
We next consider two networks consisting of three nodes and three 
edges, as shown in Fig.~\ref{fig:33}. The network motifs shown in 
Fig.~\ref{fig:33}(A) and (B) appear abundantly 
in biological networks, and they are termed 
\lq\lq feedback\rq\rq\ and \lq\lq feedforward\rq\rq\ networks, respectively \cite{milo2002network}. 
By calculating the eigenvalues and eigenvectors of the corresponding network 
Laplacians, we obtain the following expressions for $Q$ 
for Figs.~\ref{fig:33}(A) and (B):
\begin{align}
  Q^{\mathrm{(A)}} &= \frac{
  (a + b) \eta_1 + (b + c) \eta_2 + (c + a) \eta_3 }
  {18 (a b + b c + c a)},
  \label{Q33a} \\
 Q^{\mathrm{(B)}} &=
 \frac{1}{18 \left(a ^2 b + a b ^2 + a ^2 c + a c^2 + 2 a b c
 \right)} \nonumber \\
 &\qquad \left(\left(a^2 + b^2+ c^2 + 2 a b + b c \right) \eta_1
 \right. \nonumber \\
 &\qquad \  + \left(b^2 + c^2 + a b + b c + c a \right) \eta_2 \nonumber \\
 &\qquad \  \left. + \left(a ^2 + a b + a c \right) \eta_3 \right),
  \label{Q33b}
\end{align}
respectively. Because the disturbance coefficients $\alpha_i$ 
(i.e., the coefficients of $\eta_i$) are 
different for $i=1,2,3$, the $Q$ values for these cases depend on the oscillator configuration.
By restricting ourselves to the case of identical noise strengths, 
i.e., $\eta_i = \eta$ ($i=1,2,3$), we look for the optimal structures under the 
constraint $a+b+c=1$. By using, the method of Lagrange multipliers, for example, 
we find that (A) $a=b=c=\frac{1}{3}$ and (B) $a=c=\frac{1}{2}, b=0$ are 
optimal, and the corresponding $Q$ values are 
$Q^{\mathrm{(A)}}=Q^{\mathrm{(B)}}=\frac{\eta}{3}$. Thus, these two 
optimal networks are equivalently noise-tolerant. 

In network (A), even if any of 
$a,b$, or $c$ vanish, the synchronized state remains linearly stable.
However, we find that stability against noise is improved if all the connections are present. 
In contrast, the feedfoward loop in network (B) does not efficiently 
stabilize the system.
Instead, the optimal structure is a star network, in which $b$ vanishes. 


\subsection{Three oscillators with four unweighted edges} \label{sec:34}
\begin{figure}[t] 
 \includegraphics[width=50 mm]{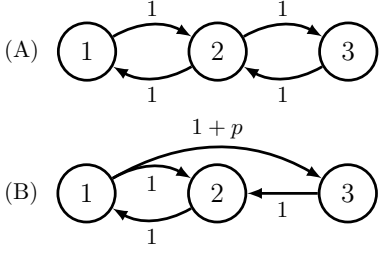}
  \caption{Networks with three nodes and four edges. Only strongly
 connected networks are considered. For $p=0$, the
 disturbance coefficients $(\alpha_1,\alpha_2,\alpha_3)$ are (a) $(5,2,5)/54$
 and (b) $(8,7,11)/144$. In both networks, the noisiest oscillator
 should be placed at node 2. For homogeneous noise strengths and $p=0$, we have
 $Q^{\rm (a)}:Q^{\rm (b)}=16:13$; thus, network (b) is more noise-tolerant than network (a).}
  \label{fig:34}
\end{figure}
We next consider networks with three nodes and four edges. 
Among such networks, we focus only on strongly connected networks, as 
shown in Fig.~\ref{fig:34}.
Instead of finding the optimal weight distribution 
for each network, we compare the $Q$ values between these two networks, 
with homogeneous weights fixed at unity. We also discuss the optimal
oscillator configuration.

For the network shown in Fig.~\ref{fig:34}(A), we obtain 
\begin{align}
  Q^{\mathrm{(A)}} &= \frac{
  5 \eta_1 + 2 \eta_2 + 5 \eta_3}
  {54}
  \label{net34a}.
\end{align}
For the network shown in Fig.~\ref{fig:34}(B), however, $L$ is not 
diagonalizable. We therefore set $A_{31}=1+p$ and calculate 
Eq.~\eqref{Q_alpha} under the assumption $p \neq 0$. As a result, we obtain 
\begin{align}
  &Q^{\mathrm{(B)}}(p)=\nonumber \\
  &\ \ \frac{(8 + 5 p + p^2) \eta_1+ (7 + 6 p + p^2)\eta_2 +  (11 + 3 p) \eta_3}
 {9(16 + 16 p + 3 p^2)}.
\end{align}
This expression is obviously continuous at $p=0$ where it reduces to
\begin{align}
  Q^{\mathrm{(B)}} &= \frac{
  8 \eta_1 + 7 \eta_2 + 11 \eta_3}
  {144}.
  \label{net34b}
\end{align}
The validity of this result is checked numerically in Sec.~\ref{sec:verification}.
Note that although we have chosen $A_{31}$ to put
an extra weight in this particular network, an extra weight to any link renders
the corresponding Laplacian diagonalizable.

When the noise strengths are homogeneous, we have 
$Q^{\mathrm{(A)}}:Q^{\mathrm{(B)}}=16:13$; thus, network (B) is 
significantly more noise-tolerant than network (A).

When the noise strengths are inhomogeneous, 
the oscillator with the largest noise strength should 
be placed at node 2 in both networks. 
One might find it reasonable because only 
node 2 has two incoming connections, whereas the other nodes each have only one.
In contrast, the difference between nodes 1 and 3 in network (B) is 
more difficult to predict. One might suppose that node 1 would disturb the network more strongly
than node 3, because nodes 1 and 3 have two and one outgoing connections,
respectively, so node 1 might have a larger $\alpha$ value. 
However, we actually have $\alpha_1:\alpha_3=8:11$; thus, node 3 
disturbs the synchronization more strongly.

%

\subsection{A ring with one directed shortcut} \label{sec:ring}
\begin{figure}[t] 
 \begin{center}
  \includegraphics[width=60 mm]{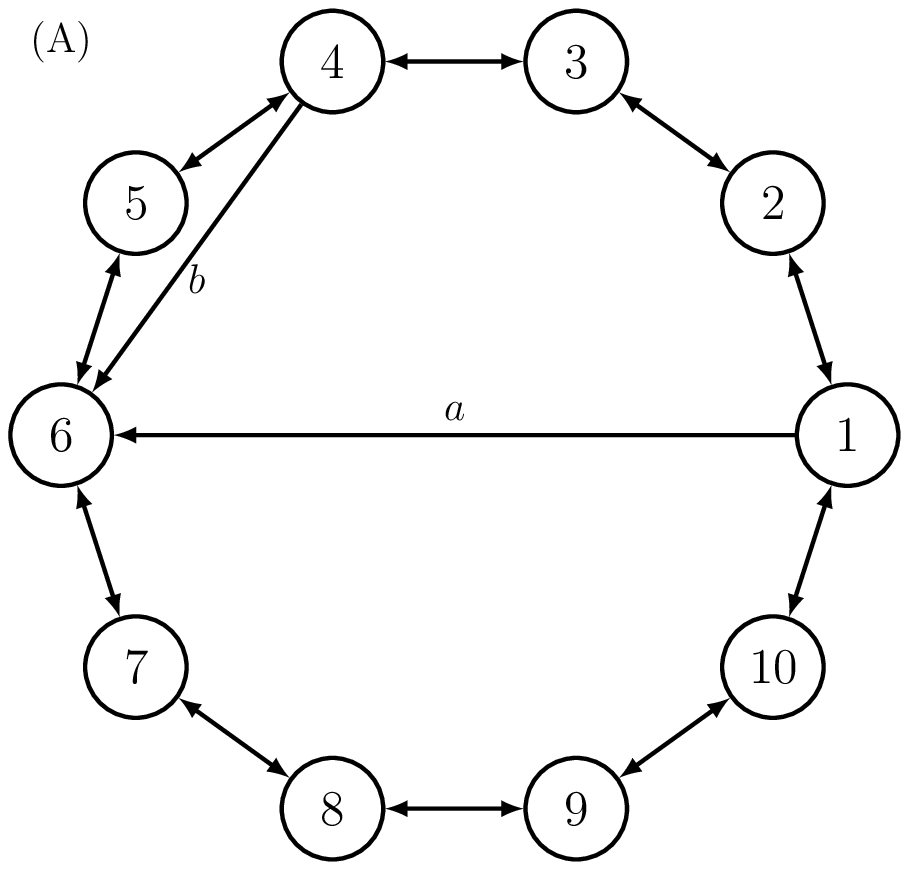}\\
  \includegraphics[width=60 mm]{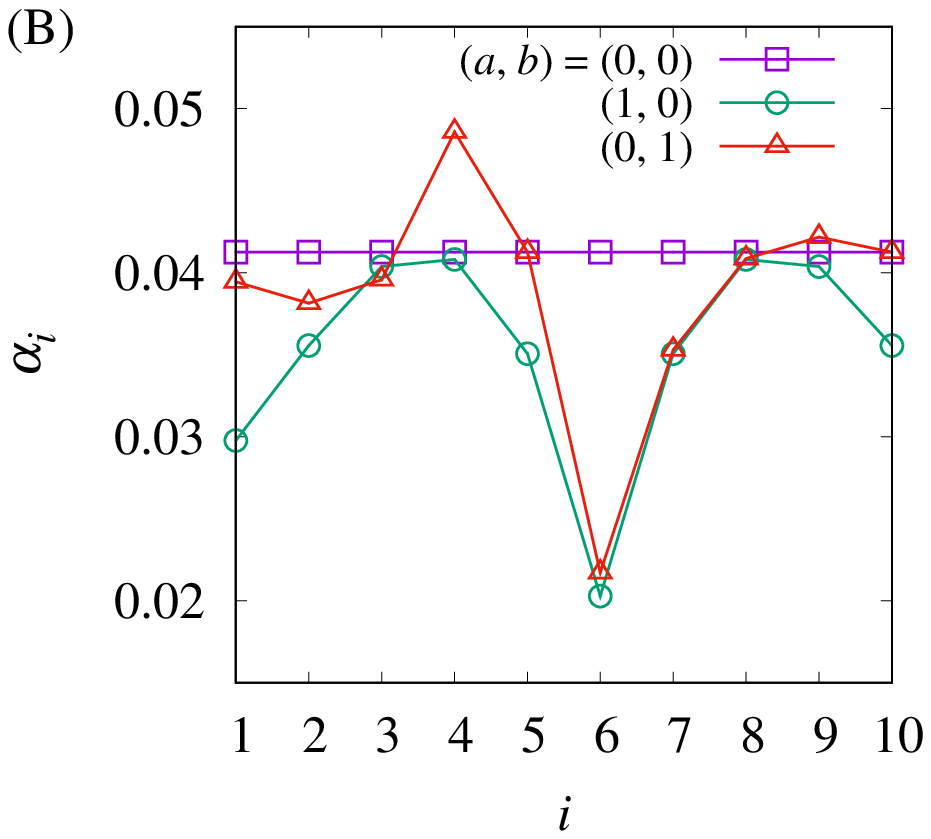}
 \end{center}
 \caption{Ring network of ten nodes with or without a shortcut. (A)
 Schematic of the network . (B) Disturbance coefficients for three cases: (i) $(a,b)=(0,0)$, (ii) $(a,b)=(1,0)$, and (iii) $(a,b)=(0,1)$.}
  \label{fig:10}
 \end{figure}
We consider the effect of a shortcut connection added 
to a network with a large path length. As depicted in 
Fig.~\ref{fig:10}(A), we consider a ring
network of ten nodes, where $A_{i, i+1}=A_{i, i-1}=1$ ($1\le i \le N$), $A_{1, N}=A_{N, 1}=1$, 
$A_{6, 1}=a, A_{6, 4}=b$, and $A_{i, j}=0$ otherwise. 
We compare three cases: (i) $(a,b)=(0,0)$, (ii) $(a,b)=(1,0)$, and (iii) $(a,b)=(0,1)$.
Figure \ref{fig:10}(b) shows the disturbance coefficients $\alpha_i$ for
the three cases. When $\eta_i=\eta$ ($1\le i
\le 10$), the corresponding $Q$ values are $Q^{\rm (i)}\simeq 0.413 \eta, Q^{\rm (ii)} \simeq 0.354 \eta$, and
$Q^{\rm (iii)} \simeq 0.388 \eta$. 
We thus find that the addition of a shortcut connection significantly 
improves the noise stability in both cases (ii) and (iii), with better 
improvement being obtained in case (ii) than in case (iii). 
We attribute the reason for this difference to the path 
length. When the path length between a pair of nodes is large, the phase 
difference between those nodes tends to be large. The shortcut connection 
in network (ii) decreases the average path length more than that of 
network (iii), resulting in better synchronization. 

Moreover, in both, cases (ii) and (iii), node 6 gets one more incoming edge. As 
shown in Fig.~\ref{fig:10}(B), this reduces the disturbance 
coefficient of node 6 considerably. Thus, when an oscillator is very noisy, its negative 
effect on synchronization can be easily suppressed by adding one 
incoming link to the oscillator. 

\subsection{A ring with frequency heterogeneity} \label{sec:ring_hetero}
\begin{figure}[t] 
 \begin{center}
  \includegraphics[width=60 mm]{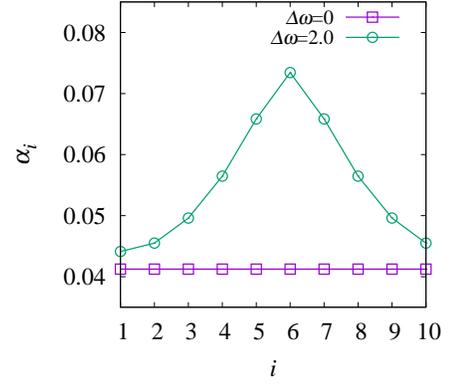}
 \end{center}
 \caption{Disturbance coefficients $\alpha_i$ of the ring network of ten nodes
 with or without frequency heterogeneity. The natural frequencies are
 $\omega_i=\omega$ for all $i$ except $\omega_6=\omega+\Delta \omega$,
 where $\omega$ is arbitrary.}
  \label{fig:10_hetero}
\end{figure}

We investigate the effect of frequency
heterogeneity using the ring network consisting of ten
oscillators, i.e., Fig.~\ref{fig:10}(A) with $a=b=0$.
We consider the case in which only one oscillator has a frequency
different from the others; i.e., $\omega_i=\omega$ for all $i$ except
$\omega_6=\omega+\Delta \omega$, where $\omega$ is arbitrary.
For this case, network Laplacian is calculated using
Eq.~\eqref{Aij_new}, where $(B_{ij})$ is the adjacency matrix for the
ring network. We assumed $f_{ij}(\cdot)=\sin(\cdot)$ and
obtained $\psi_i^*$ values ($1\le i \le N$) by simulating
Eq.~\eqref{pm2} in the absence of noise. Figure \ref{fig:10_hetero}
shows the disturbance coefficients calculated numerically using
Eq.~\eqref{alpha},
indicating that the oscillators closer to node 6 more strongly disturb synchronization.

\INS{
\subsection{A random directed network} \label{sec:random_net}
As a final example, we consider a random directed network of
100 oscillators. We employed a directed Erd\H{o}s-R\'enyi model to
generate $A$; i.e., $A_{ij} = 1$ with probability $p$ 
and $A_{ij} = 0$ otherwise for $j\neq i$; and $A_{ii} = 0$.
We set $p=0.05$, thus the mean in- and out-degrees were approximately
five in our example network.
We confirmed that the generated network suffices the stability criterion
given in Eq.~\eqref{eigenvalue} and the corresponding Laplacian is diagonalizable.
Figure \ref{fig:random_net}(A) shows the values of the disturbance
coefficients $\alpha_i$ obtained numerically using Eq.~\eqref{alpha}.
To see the relation between the values of $\alpha_i$ and the network
structure, we display two scatter plots: $\alpha_i$ vs $1/d^{\rm in}_i$
in Fig. \ref{fig:random_net}(B) and 
$\alpha_i$ vs $d^{\rm out}_i/d^{\rm in}_i$ in
Fig. \ref{fig:random_net}(C),
where $d^{\rm in}_i$ and $d^{\rm out}_i$ are
the in- and out-degrees of node $i$, respectively.
We find that
$1/d^{\rm in}_i$ is almost proportional to $\alpha_i$ and is clearly
more correlated with $\alpha_i$ than $d^{\rm out}_i/d^{\rm in}_i$.
We discuss this result later.

\begin{figure}[t] 
 \begin{center} 
  \includegraphics[width=7cm]{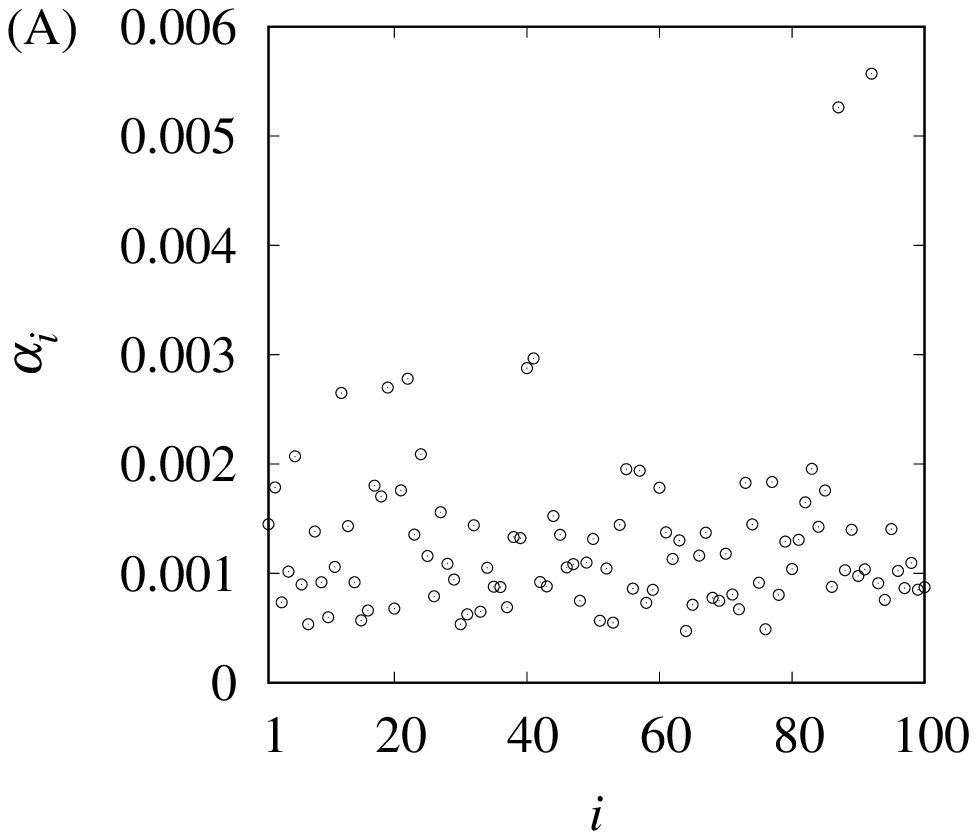} 
  \includegraphics[width=7cm]{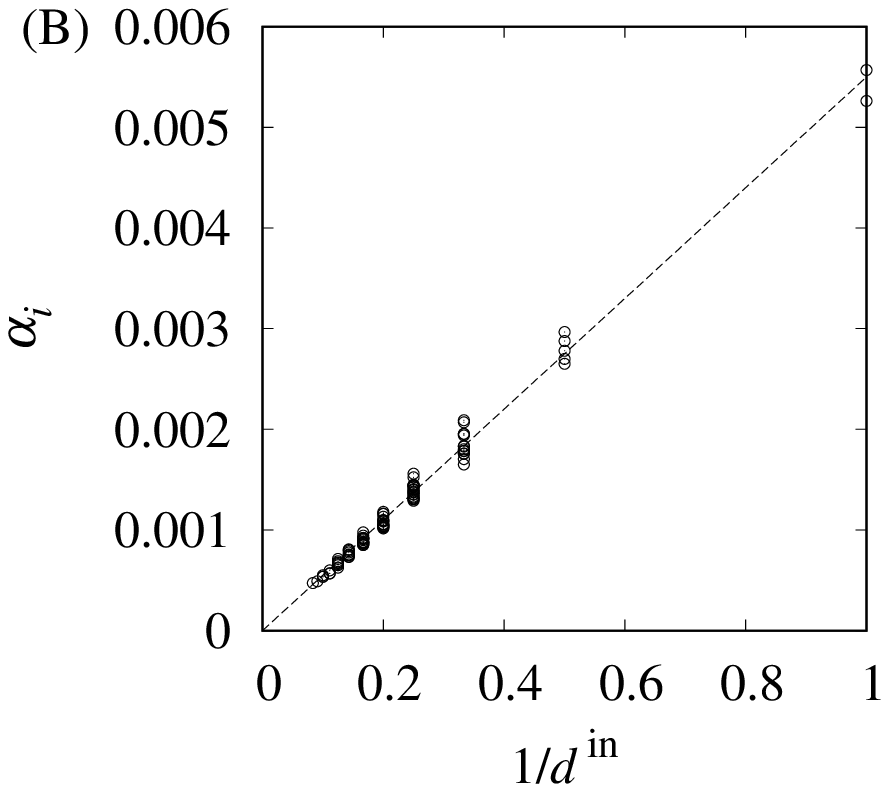} 
  \includegraphics[width=7cm]{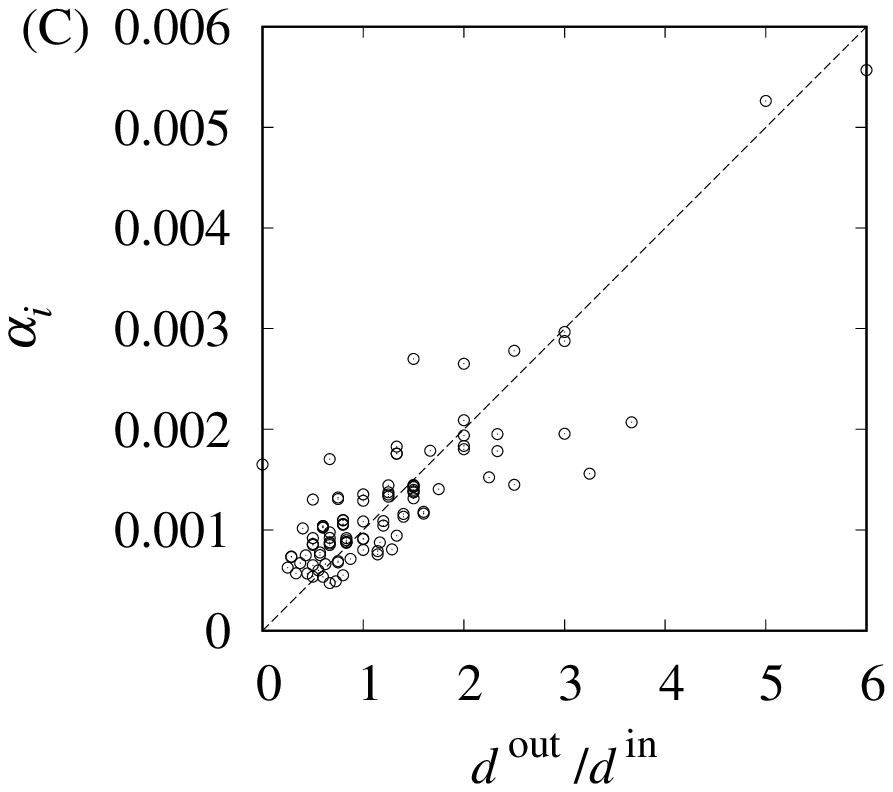} 
 \end{center} 
 \caption{A random directed network of 100  oscillators.
 (A) Values of the disturbance coefficients $\alpha_i$.
 \INS{
 (B) $\alpha_i$ vs $1/d^{\rm in}_i$. 
 (C) $\alpha_i$ vs $d^{\rm out}_i/d^{\rm in}_i$.
 Dashed lines are for the guidance to eye, with slopes 0.0055 and 0.001
 in (B) and (C), respectively.
 }
 }
 \label{fig:random_net}
\end{figure}
}

\section{Numerical verification} \label{sec:verification}
\begin{figure}[t] 
 \begin{center} 
  \includegraphics[width=7cm]{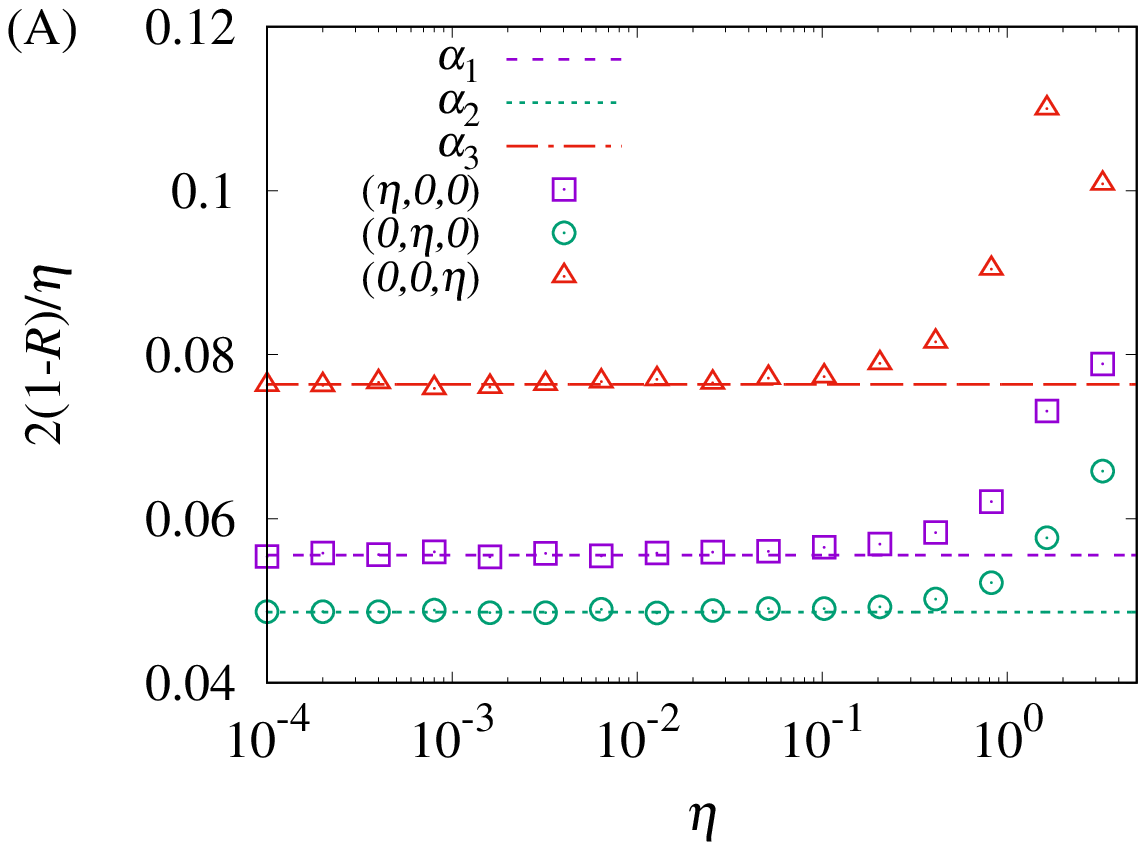}
  \includegraphics[width=7cm]{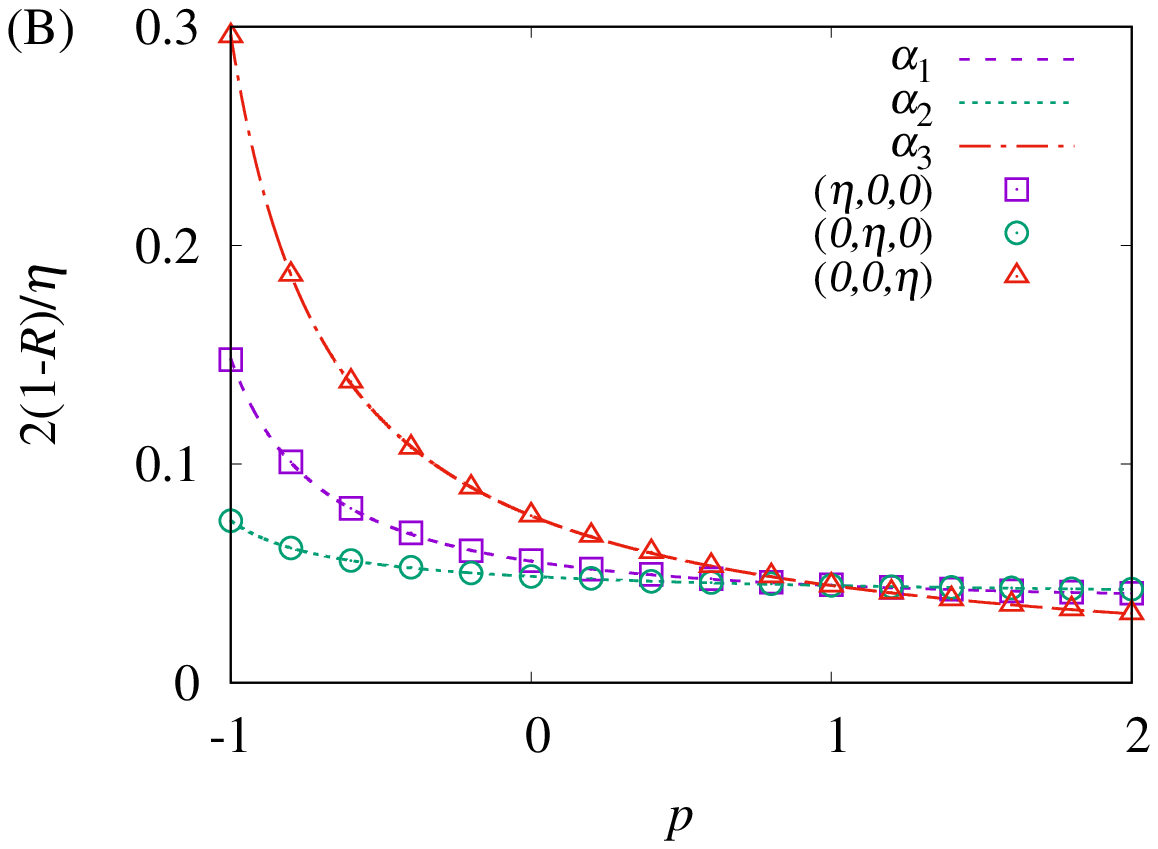} 
  \includegraphics[width=7cm]{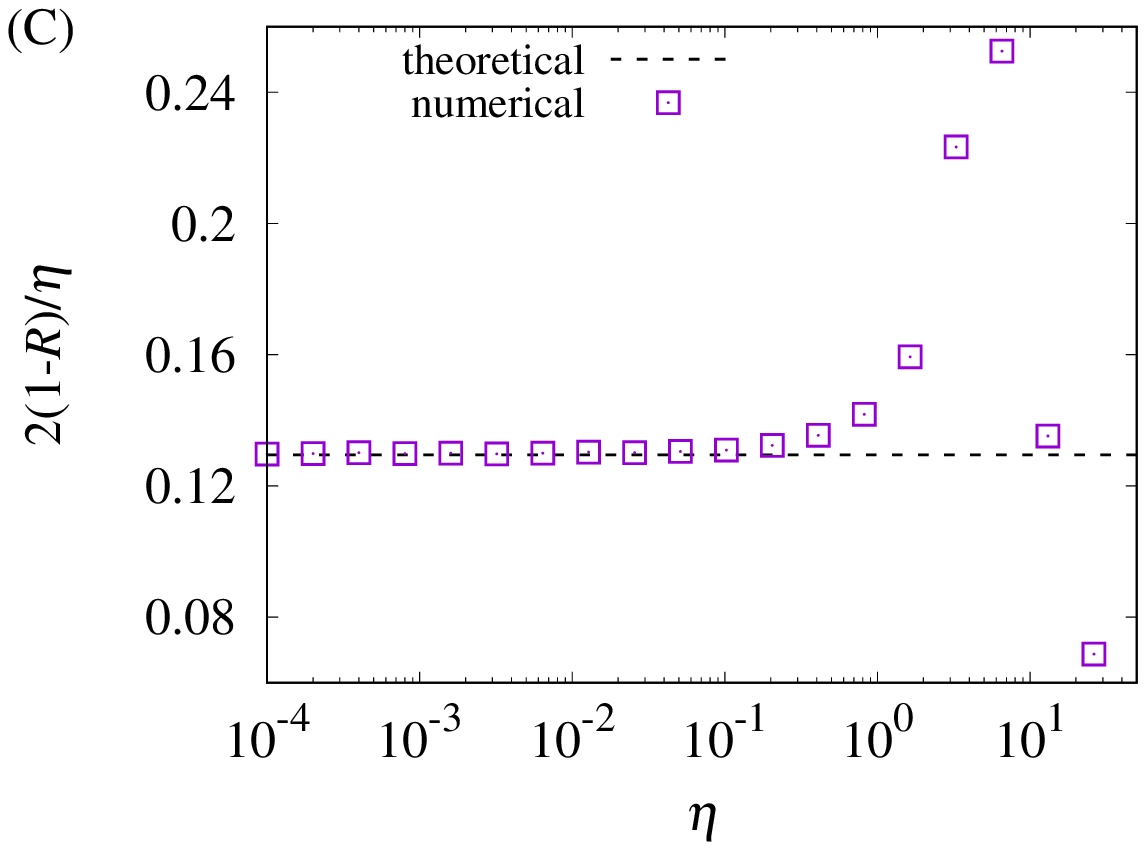} 
 \end{center} 
 \caption{Numerical verification with example networks. (A) Values of $2(1-R)/\eta$ for different 
 noise strengths $\eta$ for the network in Fig.~\ref{fig:34} with
 $p=0$. (B) Values of $2(1-R)/\eta$ for the network in Fig.~\ref{fig:34}
 with different $p$ values, where $\eta=0.01$.
 (C) Values of $2(1-R)/\eta$ for the directed random network used in \ref{sec:random_net}.
 Here, $R$ is the long-time-averaged Kuramoto order parameter, 
 which is obtained from the numerical simulations of Eq.~\eqref{pm}. 
 The numerical values are in excellent agreement with theoretical 
 predictions.}
 \label{fig:num}
\end{figure}

Using the example network shown in Fig.~\ref{fig:34}(B), we have 
verified our theory numerically. We simulated Eq.~\eqref{pm} numerically with
$f(\cdot) = \sin(\cdot)$ using random initial conditions, and we
measured the Kuramoto order parameter $r(t) = 
\frac{1}{N}\left|\sum_{j=1}^N e^{i\phi_j} \right|$. 
The long-time average of $r(t)$, denoted by $R$, is expected to provide a good 
approximation to $\langle r \rangle$. In our simulations, we measured 
\begin{equation}
 R = \frac{1}{t_1-t_0}\int_{t_0}^{t_1} r(t) dt,
  \label{R}
\end{equation} 
where $t_0=1000$ and $t_1=10000$. 
Furthermore, from Eqs.~\eqref{r_expectation} and \eqref{Q_alpha}, it 
follows that $Q=\sum_{i} \alpha_i \eta_i = 2(1-\langle r \rangle)$. 
Thus, by setting $(\eta_1,\eta_2,\eta_3) = (\eta,0,0), (0,\eta,0)$, or 
$(0,0,\eta)$, we expect the quantity $2(1-R)/\eta$ to coincide with $\alpha_i$ 
$(i=1,2,3)$, respectively. 
In Fig.~\ref{fig:num}(a), we plot the values of $2(1-R)/\eta$ for different 
values of $\eta$. For small $\eta$, the numerical data are in excellent agreement with the theoretically predicted $\alpha_i$ values. 
However, for large $\eta$, there are considerable deviations, which 
are due to the nonlinear effects in our model. 

As mentioned earlier, the network shown in Fig.~\ref{fig:34}(B) for 
$p=0$ yields a 
non-diagonalizable Laplacian $L$. 
We have measured the values of $2(1-R)/\eta$ numerically for different $p$ values, 
as shown in Fig.~\ref{fig:num}(B). The numerical values of $2(1-R)/\eta$ are in 
excellent agreement with the theoretical values of the $\alpha_i$, even for $p=0$, 
at which point $L$ becomes non-diagonalizable. This result supports the 
validity of the method proposed for treating non-diagonalizable matrices $L$ in 
Sec.~\ref{sec:nondiagonalizable}. 

We then performed numerical simulation of Eq.~\eqref{pm}
for the directed random network employed in \ref{sec:random_net}
with homogeneous noise strength $\eta_i = \eta$.
In this case, $Q=\sum_i \alpha_i \eta \simeq 12.9 \eta$.
As shown in Fig.~\ref{fig:num}(C), simulation data and the predicted
$Q$ value are in excellent agreement for small $\eta$.


\section{Discussion and conclusions} \label{sec:conclusion} 
We have provided a theoretical framework for quantifying the magnitude $Q$ 
of the fluctuations around the synchronous state of a given oscillator network.
We have also provided several example networks to discuss the optimal or better
network structures. 
Given a nonlinear dynamical system or a network Laplacian, its $Q$ value 
is readily computable. Using these $Q$ values, we can quantitatively compare 
the noise stability of the networks 
of different numbers of 
nodes and edges with possibly heterogeneous, signed weights.
Furthermore, the disturbance coefficients $\alpha_i$, which 
appear in the expression for $Q$, 
represent how strongly an oscillator at node $i$ disturbs synchronization. 
Using the values of $Q$ and $\alpha_i$, we can find the optimal network structure 
and the optimal oscillator configuration, as demonstrated in 
Sec.~\ref{sec:examples}. 

In the example shown in Fig.~\ref{fig:10}, we show that shortcut
connections are effective for making oscillator networks
noise-tolerant. Such networks are often referred to as small-world
networks \cite{watts98}, and there is a large body of
theoretical results indicating that synchronization is enhanced as
the number of shortcuts increases.
Among them, the study by Korniss et al. \cite{korniss03}
is very relevant to the present study. They employed a 
course-grained description of the oscillator network to show that
shortcut connections added to lattice networks
prevent the divergence of the phase variance, given in Eq.~\eqref{Q},
as $N$ goes infinity \cite{korniss03}.
Such an approach is certainly powerful for understanding typical properties
shared by certain network classes.
Our approach can be regarded as a complementary one. We
can quantify fluctuations in synchronized dynamics
in particular networks of any class in a detailed manner.

Our study is based on a general class of linear dynamical systems 
with additive noise, given in Eq.~\eqref{model}. 
There are other theoretical studies concerning the same linear systems that 
treat different quantities of interest. 
For example, Refs. \cite{kori09,mkk10,cross2012improving} investigate
the dynamics of the collective mode of an oscillator 
network. This problem can concisely be formulated as a projection 
of the entire dynamical system onto a one-dimensional dynamical mode along 
the synchronization manifold, which is $\bm u^{(1)}$ in the present theory. 
For example, when oscillators are subjected to 
independent noise, as we consider in the present paper, 
the diffusion coefficient of the collective mode can 
be derived as a function of $\bm v^{(1)}$ \cite{mkk10}. Moreover, it has been shown that 
element $w_i$ of $\bm v^{(1)}=(w_1\, w_2\, \ldots\, w_N)$ describes 
the strength of the influence of node $i$ on the collective mode
\cite{kori09,mkk09pre,skardal2016collective}.

\INS{
We emphasize that $w_i$ and $\alpha_i$ are different measures because they are
related to the dynamics along and transverse to the synchronization
manifold, respectively. Therefore, they are not necessarily correlated.
For example, for symmetric $L$, $w_i$ is constant for all nodes whereas
$\alpha_i$ can be heterogeneous. Actually, as shown in
Fig. \ref{fig:10_hetero}, $\alpha_i$ is heterogeneous for $\Delta
\omega=2.0$ in spite of symmetric $L$. However, in large directed random
networks, they seem to be positively correlated because $w_i$ is roughly
proportional to $d^{\rm out}_i/d^{\rm in}_i$, which is derived using a
mean-field approximation \cite{mkk09njp}, whereas $\alpha_i$ is
approximately proportional to $1/d^{\rm in}_i$ as is numerically found
in Fig.~\ref{fig:random_net}(B).  Namely, a node with a small incoming
degree tends to have large $w_i$ and $\alpha_i$ values.  The property
$\alpha_i \sim 1/d^{\rm in}_i$ is not theoretically rationalized and
remains an important open problem.  However, it makes sense that
$\alpha_i$ tends to be larger for smaller $d^{\rm in}_i$ because such
nodes can only weakly tune their own rhythm to others and thus more
strongly disturb the population.
}

Ref. \cite{mkk12} treats the precision of the cycle-to-cycle periods 
of a synchronous state in an oscillator network. This 
problem involves all the dynamical modes, as is also the case for the present 
problem. However, the major contribution to the fluctuations in cycle-to-cycle 
periods comes from the dynamical mode along the synchronization 
manifold; in contrast, our problem is independent of such a mode. This is the 
reason why the contribution of the zero eigenmode is absent from our 
expression for $Q$; i.e., the summation in Eq.~\eqref{Q} starts from 
$m,n=2$.

\INS{
Many studies on the stability of synchronization focus on a few
eigenmodes, such as the mode associated with $\lambda_2$ because it
characterizes the long-time behavior of the relaxation process to a
synchronized state in the absence of noise.  In contrast, when noise
is present, noise keeps to excite all the eigenmodes. Noise stability
is thus involved with all the eigenmodes, as reflected in the
expressions for $Q$ and $\alpha_i$. When a part of eigenvalues have
vanishingly small real parts, the contributions of other eigenmodes
can be neglected in those expressions.  However, such a situation is
exceptional, such as when the system is near the
synchronization-desynchronization transition point.
}

Synchronization is essential in 
various artistic performances, including those of orchestras, choruses, and dancers. 
To improve synchronization in such performances, 
our theory may be helpful in indicating a better network structure, 
the placement of experts and laymen, and who to have look at whom. 
Experimental study, such as synchronization continuation of finger
tapping \cite{okano2017paired}, is required to demonstrate our theoretical study.

\section*{Acknowledgments}
This work was motivated by the discussion with Dr. Manabu Honda
(National Center of Neurology and Psychiatry) about kecak,
a form of Balinese dance and music drama in Indonesia. 
This work was supported by 
MEXT KAKENHI Grant Number 15H05876 (Non-linear Neuro-oscillology)
and JSPS KAKENHI Grant Number 18K11464.

%

\onecolumngrid
\appendix
\section{Derivation of Eq.~\eqref{Q_alpha}}\label{sec:derivation}
We decompose $\bm x$ as
\begin{equation}
 \bm x(t) = \sum_{m=1}^N y_m(t) \bm u^{(m)},
  \label{x_decompose}
\end{equation}  
where $ y_m (t)$ is given by
\begin{align}
 y_m (t) = {\bm v}^{(m)} \bm x (t).
 \label{y_m}
\end{align}
By taking the time derivative of Eq.~\eqref{y_m}
and using Eqs.~\eqref{model_x} and \eqref{v}, we obtain
\begin{align}
 \dot y_m (t)
  = -  \lambda_m y_m (t)
  + \hat \xi_m (t),
 \label{dot_y}
\end{align}
where
\begin{align}
  \hat \xi_m (t)
  = \sum _{i = 1}^{N} v_{i}^{(m)} \xi_i (t).
 \label{hat_xi_m}
\end{align}
It is straightforward to show that
\begin{equation}
 \langle \hat \xi_m (t) \rangle = 0, \quad
  \langle \hat \xi_m (t) \hat \xi_n (s) \rangle = \hat \eta_{mn} \delta
  (t - s),
  \label{noise2}
\end{equation}
where
\begin{align}
 \hat \eta_{mn} = \sum _{i = 1}^{N}
 v_{i}^{(m)} v_{i}^{(n)} \eta_i.
 \label{hat_eta_mn} 
\end{align}
The solution to Eq.~\eqref{dot_y} can be formally written as
\begin{align}
 y_m (t) = e^{-  \lambda_m t} y_m (0)
 + \int _{0}^{t} e^{-  \lambda_m (t - s)} \hat \xi_m (s) ds.
 \label{ym_formal}
\end{align}
For $m,n \ge 2$, using Eqs.~\eqref{noise2} and \eqref{ym_formal}, we obtain
\begin{align}
 &\left\langle y_m (t) y_n (t) \right\rangle \nonumber \\
&=\left\langle \left(e^{-  \lambda_m t} y_m (0)
 + \int _{0}^{t} e^{-  \lambda_m (t - s_1)} \hat \xi_m (s_1) ds_1\right)
 \left(e^{-  \lambda_n t} y_n (0)
 + \int _{0}^{t} e^{-  \lambda_n (t - s_2)} \hat \xi_n (s_2) ds_2\right)
 \right\rangle \\
&=\left\langle e^{-  \lambda_m t} y_m (0) e^{-  \lambda_n t} y_n (0)
 \right\rangle 
+ \left\langle e^{-  \lambda_m t} y_m (0)
 \int _{0}^{t} e^{-  \lambda_n (t - s_2)} \hat \xi_n (s_2) ds_2
 \right\rangle 
+ \left\langle e^{-  \lambda_n t} y_n (0)
 \int _{0}^{t} e^{-  \lambda_m (t - s_1)} \hat \xi_m (s_1) ds_1
 \right\rangle \nonumber\\
& \qquad + \left\langle 
 \int _{0}^{t} e^{-  \lambda_m (t - s_1)} \hat \xi_m (s_1) ds_1
 \int _{0}^{t} e^{-  \lambda_n (t - s_2)} \hat \xi_n (s_2) ds_2
 \right\rangle \\
 &= e^{-  (\lambda_m+\lambda_n) t} y_m (0) y_n (0)
 + \left\langle  \int _{0}^{t} ds_1 \int _{0}^{t} ds_2
 e^{-  \lambda_m (t - s_1)} \hat \xi_m (s_1) 
  e^{-    \lambda_n (t - s_2)} \hat \xi_n (s_2)  \right\rangle \\
&= e^{-  (\lambda_m+\lambda_n) t} y_m (0) y_n (0)
 + \int _{0}^{t} ds_1 \int _{0}^{t} ds_2
 e^{-  \lambda_m (t - s_1)} e^{-  \lambda_n (t - s_2)} 
 \left\langle \hat \xi_m (s_1)  \hat \xi_n (s_2) \right\rangle \\
&= e^{-  (\lambda_m+\lambda_n) t} y_m (0) y_n (0)
 + \int _{0}^{t} ds_1 \int _{0}^{t} ds_2
 e^{-  \lambda_m (t - s_1)} e^{-  \lambda_n (t - s_2)} 
 \hat \eta_{mn} \delta (s_1 - s_2) \\
&= e^{-  (\lambda_m+\lambda_n) t} y_m (0) y_n (0)
 + \hat \eta_{mn} \int _{0}^{t} ds
 e^{-  (\lambda_m + \lambda_n) (t - s)} \\
&= e^{-  (\lambda_m+\lambda_n) t} y_m (0) y_n (0)
 + \hat \eta_{mn} \frac{ 1 - e^{-  (\lambda_m+\lambda_n) t}}{\lambda_m + \lambda_n}\\
&\to \frac{\hat \eta_{mn} }{\lambda_m + \lambda_n} \quad (t\to \infty)
\end{align}
Here, we take the limit $t\to\infty$ because we are interested in a steady
process in which the dependence on initial conditions vanishes.

Now we derive the expression for $Q$.  
For convenience, we rewrite the definitions:
\begin{align}
 &\overline x = \frac{1}{N} \sum_{j=1}^N x_i,\\
 &\overline{u ^{(m)}} = \frac{1}{N} \sum_{j=1}^N u_{j}^{(m)},\\
 &\overline{ u^{(m)} u^{(n)} } = \frac{1}{N}
 \sum_{j=1}^N u_{j}^{(m)} u_{j}^{(n)}
\end{align}
Using Eq.~\eqref{x_decompose}, i.e., $x_j = \sum _{m = 1}^{N} u_{j}^{(m)} y_m$,
we obtain
\begin{align}
 Q &=\frac {1}{N} \sum _{j = 1}^{N} \left\langle (x_j - \overline x)^2 \right\rangle\\
   &= \frac {1}{N} \sum _{j = 1}^{N} \left\langle
   \left(\sum _{m = 1}^{N}
   \left(u_{j}^{(m)} - \overline {u ^{(m)}}
   \right) y_m \right)
   \left(\sum _{n = 1}^{N}
   \left(u_{j}^{(n)} - \overline {u ^{(n)}}
   \right) y_n \right)
   \right\rangle
   \nonumber \\
   &= \frac {1}{N} \sum _{j = 1}^{N}\sum _{m, n = 1}^{N}
   \left(
   u_{j}^{(m)} u_{j}^{(n)} - u_{j}^{(m)} \overline {u ^{(n)}}
   - \overline {u ^{(m)}} u_{j}^{(n)} + \overline {u ^{(m)}} \ \overline {u ^{(n)}}
   \right)
   \left\langle
   y_m (t) y_n (t)
   \right\rangle.\\
  &= \sum _{m,n=1}^{N}
   \left(
   \overline {u ^{(m)}  u ^{(n)}} - \overline {u ^{(m)}} \ \overline {u
   ^{(n)}}
   \right)
   \left\langle y_m (t) y_n (t) \right\rangle \label{Q1}\\
  &= \sum _{m,n=2}^{N}
   \left(
   \overline {u ^{(m)}  u ^{(n)}} - \overline {u ^{(m)}} \ \overline {u
   ^{(n)}}
   \right)
   \left\langle y_m (t) y_n (t) \right\rangle \label{Q2}\\
  &= \sum _{m,n= 2}^{N}
   \left(
   \overline {u ^{(m)}  u ^{(n)}} - \overline {u ^{(m)}} \ \overline {u
   ^{(n)}}
   \right)
   \frac {\hat \eta_{mn}}{\lambda_m + \lambda_n},\\
  &= \sum _{m,n= 2}^{N}
   \left(
   \overline {u ^{(m)}  u ^{(n)}} - \overline {u ^{(m)}} \ \overline {u
   ^{(n)}}
   \right)
 \frac {\hat \eta_{mn}}{ \lambda_m + \lambda_n},\\
  &= \sum _{i = 1}^{N} \sum _{m,n= 2}^{N}
   \frac{
   \overline {u ^{(m)}  u ^{(n)}} - \overline {u ^{(m)}} \ \overline {u
 ^{(n)}}}
 { \lambda_m + \lambda_n}
  {v_{i}^{(m)} v_{i}^{(n)}} \eta_i,
\end{align}
which is Eq.~\eqref{Q_alpha}.
To pass from Eq.~\eqref{Q1} to Eq.~\eqref{Q2}, we have used the relation
\begin{equation}
 \overline {u ^{(m)}  u ^{(n)}} - \overline {u ^{(m)}}\  \overline
  {u^{(n)}} = 0 \quad \mbox{for $m=1$ or $n=1$},
\end{equation}
which holds because $\bm u^{(1)}=(1,1,\ldots,1)^{\rm T}$.
%

For a symmetric matrix $L$,
Eq.~\eqref{alpha} reduces to Eq.~\eqref{alpha_sym}
because $\bm v ^{(n)} = (\bm u ^{(n)})^{\mathrm{T}}$,
$\bm u ^{(m)} \cdot \bm u ^{(n)} = \delta _{mn}$
for $1 \leq m, n \leq N$, 
$\overline{u^{(n)}} = \frac {1}{N} \sum _{i = 1} ^{N} u_{i}^{(n)}
\propto \sum _{i = 1} ^{N} u_{i}^{(1)} u_{i}^{(n)}
= \bm u ^{(1)} \cdot \bm u ^{(n)} = 0$, and
   $\overline {u^{(m)}  u^{(n)}}
   = \frac {1}{N} \, \bm u ^{(m)} \cdot \bm u ^{(n)}
   = \frac {\delta _{mn}}{N}$.

\end{document}